\documentclass[aps,prd,preprint,a4paper,showpacs]{revtex4}
\usepackage{graphicx,natbib}
\usepackage{amsmath}

\begin{document}
\title{Neutrino Large Mixing in Universal Yukawa Coupling Model with Small 
Violation}
%\date{\today}
\vspace*{1cm}
\author{T. Teshima}
\email{teshima@isc.chubu.ac.jp}
\author{T. Asai}
\author{Y. Abe} 
\affiliation{Department of Applied Physics,  Chubu University, Kasugai 
487-8501, Japan}
\begin{abstract}
We have analyzed the possibility that the universal Yukawa coupling 
(democratic mass matrix) with small violations of Dirac and Majorana neutrinos 
can induce the large mixing of neutrinos through the seesaw mechanism. The 
possibility can be achieved by the condition that the violation parameters of 
Majorana neutrinos are sufficiently smaller than the violation parameters of 
Dirac neutrinos. Allowed regions of the violation parameters producing the 
observed neutrino mass hierarchy and large neutrino mixing are not so 
restricted at present in contrast to the violation parameters for quark 
sector. In order to obtain the distinctive features in violation parameters 
for neutrino sector as quark sector, one should take more precise data 
concerned with the neutrino experiments, for example  mass values of neutrinos 
, mixing parameters of $U_{\rm MNS}$ and CP violation phases.
\end{abstract}
\pacs{12.15.Ff, 13.15.+g, 14.60.Pq}
\preprint{CU-TP/02-06}
\maketitle
%\setlength{\baselineskip}{0.33in}
%%%%%%%%%%%%% section 1%%%%%%%%%%%%%%%
\section{Introduction}
Recent neutrino experiment in Super-Kamiokande \cite{SUPERKAMIOKANDEI} on
atmospheric neutrinos 
has confirmed the $\nu_\mu\leftrightarrow \nu_\tau$ oscillation being 
large mixing, $\sin^22\theta_{\rm atm}>0.88$, and found the range of the mass 
parameter $\Delta m_{\rm atm}^2$ to be $(1.5-5)\times10^{-3}~{\rm eV}^2$. 
Solar neutrino experiments by the Super-Kamiokande collaboration 
\cite{SUPERKAMIOKANDEII} favor the large mixing angle (LMA) MSW solution, 
although there can be four solutions, the small mixing angle (SMA) MSW 
solution, the low $\Delta m^2$ (LOW) solution and the vacuum (VAC) solution.  
The mass and mixing parameter $(\Delta m^2_{\odot},\ \sin^22\theta_\odot)$ 
for LMA-MSW solution is $(3.2\times10^{-5}{\rm eV^2},\ 0.75)$ 
\cite{SUPERKAMIOKANDEII}. 
As we consider the three-flavor neutrinos, we can set $\Delta m^2_{\rm atm}$ 
to $\Delta m^2_{23}$ and $\Delta m^2_{\odot}$ to $\Delta m^2_{12}$, and 
$\sin^22\theta_{\rm atm}=\sin^22\theta_{23}\sim1$ and $\sin^22\theta_{\odot}=
\sin^22\theta_{12}\sim1$. The remaining mixing angle $\theta_{13}$ is 
restricted to $\sin^22\theta_{13}<0.10$ by the CHOOZ experiment \cite{CHOOZ}. 
\par
To answer a question why the neutrino sector mixings expressed by a 
$U_{\rm MNS}$ matrix \cite{MNS} are so large in contrast to the small mixings 
of the quark sector expressed by a $U_{\rm CKM}$ matrix is one of the most 
challenging issues at present. Furthermore, the values of neutrino masses 
predicted in neutrino oscillation lead to the next question why the neutrino 
masses are so small compared to other Fermion masses. In order to explain 
the second question, almost authors assume the seesaw mechanism \cite{SEESAW} 
introducing the heavy right handed Majorana neutrinos. For the explanation of 
first questions, there are two scenarios; first \cite{MODELS,KING} is to use 
the so-called Froggatt Nielsen mechanism \cite{FNM} in order to make the mass 
hierarchy of lepton masses, and second \cite{YANAGIDA,BRANCO,TESHIMA1} to use 
the democratic mass matrix with small violations in order to make the mass 
hierarchy of lepton masses. We call the second type of mass matrix "universal 
type mass matrix" hereafter. In the first scenario, the matrix elements except 
33 element of mass matrix are suppressed by the powers of spontaneous 
breaking of some family symmetry. Then, we call the first type of mass matrix 
"family symmetry dependent type mass matrix". 
\par 
If one assumes a standpoint of grand unified theory (GUT), the mass matrices  
of quarks and charged leptons should be the same type. 
Mass matrix for Dirac neutrino in models should have the same type of quark 
and charged lepton mass matrix. Furthermore, it is natural to assume that 
the mass matrix of the Majorana neutrino is the same type of mass 
matrix for other Dirac Fermion. In scenarios assuming the family symmetry 
dependent type mass matrix, a few models \cite{KING} adopt the mass matrix of 
the Majorana neutrino similar to the type of mass matrix for other Dirac 
Fermions. In models assuming the universal type mass matrix, literature 
\cite{BRANCO,TESHIMA1} adopts the mass matrix for the Majorana neutrino 
similar to the type of mass matrix for other Dirac Fermion.  
\par 
Works in Refs.~\cite{BRANCO,TESHIMA1} use the universal Yukawa coupling 
(democratic mass matrix) with small violations for Dirac neutrino and also for 
Majorana neutrino, and discuss the possibility producing the neutrino large 
mixing through the seesaw mechanism. We call this possibility "neutrino large 
mixing induced through seesaw mechanism". Neutrino mixing matrix 
$U_{\rm MNS}$ is 
given by a product of an unitary matrix $U_l$ diagonalizing the charged 
lepton mass and a unitary matrix $U_{\nu}$ diagonalizing the effective 
neutrino mass $M_DM_M^{-1}M_D^{T}$, where $M_D$ and $M_M$ are the Dirac 
neutrino and the Majorana neutrino mass matrix, respectively. 
The charged lepton mass matrix is the democratic mass matrix 
$M_0=m{\footnotesize \left( \begin{array}{ccc}
                      1&1&1\\
                      1&1&1\\
                      1&1&1
               \end{array}\right)}$ 
in the limit neglecting small violations, and this mass matrix is diagonalized 
by the unitary matrix 
$U_0={\footnotesize \left( \begin{array}{ccc}
                      1/\sqrt{2}&-1/\sqrt{2}&0\\
                      1/\sqrt{6}&1/\sqrt{6}&-2/\sqrt{6}\\
                      1/\sqrt{3}&1/\sqrt{3}&1/\sqrt{3}
               \end{array}\right)}$ 
in the limit neglecting small violations. For the neutrino masses, the 
effective neutrino mass $M_{\rm eff}$ is produced through the seesaw mechanism 
as $M_{\rm eff}=M_DM^{-1}_MM_D^t$. Though these neutrino mass 
matrices $M_D$ and $M_M$ are democratic type mass matrices, the effective 
neutrino mass matrix $M_{\rm eff}$ could be almost diagonal if the small 
violations in $M_M$ satisfy a certain condition as discussed in section III. 
If the effective neutrino mass matrix $M_{\rm eff}$ is almost diagonal, the 
transforming matrix $U_{\nu}$ for the neutrino is close to the unit matrix. 
Then the neutrino mixing matrix $U_{\rm MNS}$ satisfies $U_{\rm MNS}\sim U_0$, 
and large neutrino mixing is realized. 
%%%%%%%%%%% section 2 %%%%%%%%%%%%%%
\section{Neutrino Large Mixing and Violation of universal Yukawa coupling}
In the universal Yukawa coupling scenario, the main mass hierarchy is 
produced by the universality of the Yukawa coupling (democratic mass matrix 
\cite{UNIVERSAL}), 
and another mass hierarchy is produced by small violations added to the 
democratic mass matrix. This violation is considered to be just like the 
$SU(3)$ violation in the hadron spectroscopy and hadron decay processes. This 
$SU(3)$ violation is considered to be produced by the quark mass difference 
(violation from the $SU(3)$ symmetry) and quark dynamics. Similarly, the 
violations added to the democratic mass matrix are considered to be produced 
by some violation from a horizontal symmetry and some dynamics of the quarks 
and leptons. Because the origin of the violation is not clear at present, we 
treat these small violations as free parameters. 
\par
We assume the following mass matrices for charged leptons and Dirac 
neutrinos;  
\begin{eqnarray}
&&M_l=\Gamma_l\left(\begin{array}{ccc}
                1&1-\delta^l_1&1-\delta^l_2\\
                1-\delta^l_1&1&1-\delta^l_3\\
                1-\delta^l_2&1-\delta^l_3&1
                \end{array}\right),\ \ \ 
\delta^l_i\ll1\ \  (i=1,2,3),\\
&&M^{\nu}_D=\Gamma^{\nu}_D\left(\begin{array}{ccc}
                1&1-\delta^{\nu}_1&1-\delta^{\nu}_2\\
                1-\delta^{\nu}_1&1&1-\delta^{\nu}_3\\
                1-\delta^{\nu}_2&1-\delta^{\nu}_3&1
                \end{array}\right),\ \ \ 
\delta^{\nu}_i\ll1\ \  (i=1,2,3).
\end{eqnarray}
This violation pattern is similar to that of the quark sector(neglecting 
the phases) \cite{TESHIMA2}, 
\begin{equation}
M_q=\Gamma_q\left(\begin{array}{ccc}
                1&1-\delta^q_1&1-\delta^q_2\\
                1-\delta^q_1&1&1-\delta^q_3\\
                1-\delta^q_2&1-\delta^q_3&1
                \end{array}\right),\ \ \ 
\delta^q_i\ll1\ \  (q=u,d,\ i=1,2,3).
\end{equation}
These mass matrices are diagonalized by a unitary matrix $U(\delta_1, 
\delta_2, \delta_3)$ close to $U_0$, i.e., 
\begin{equation}
U(\delta_1, 
\delta_2, \delta_3)\sim U_0= \left( \begin{array}{ccc}
                      1/\sqrt{2}&-1/\sqrt{2}&0\\
                      1/\sqrt{6}&1/\sqrt{6}&-2/\sqrt{6}\\
                      1/\sqrt{3}&1/\sqrt{3}&1/\sqrt{3}
               \end{array}\right).
\end{equation}
Eigenvalues are obtained as 
\begin{eqnarray}
&&m_1\approx \left[\frac13(\delta_1+\delta_2+\delta_3)-\frac13
      \xi\right]\Gamma\approx \delta_1\Gamma,
      \nonumber\\
&&m_2\approx \left[\frac13(\delta_1+\delta_2+\delta_3)+\frac13
      \xi\right]\Gamma\approx\frac23(\delta_2+\delta_3)\Gamma,\\
&&m_3\approx \left[3-\frac23(\delta_1+\delta_2+\delta_3)
      \right]\Gamma\approx3\Gamma,\nonumber
\end{eqnarray}  
where
\begin{equation}
\xi=\left[(\delta_2+\delta_3-2\delta_1)^2+3(\delta_2-\delta_3)^2
      \right]^{1/2}.
\end{equation} 
In the same democratic mass matrix scenario, Branco et al.~\cite{BRANCO} 
assumed the following pattern of violations for charged leptons $l$ and Dirac 
neutrinos $D$. 
\begin{equation}
M_i=\Gamma_i\left(\begin{array}{ccc}
                1&1&1\\
                1&1+\varepsilon^i_1&1\\
                1&1&1+\varepsilon^i_2
                \end{array}\right),\ \ \ 
\varepsilon^i_1,\ \varepsilon^i_2\ll1\ \  (i=l, D).
\end{equation}
\par
For neutrino masses, we use the seesaw mechanism. Then the effective 
left-handed neutrino masses are given as 
\begin{equation}
M^{\nu}_{\rm eff}=M^{\nu}_D{M^{\nu}_M}^{-1}(M^{\nu}_D)^t,
\end{equation}
using the right-handed Majorana neutrino mass matrix $M^{\nu}_M$. For this 
Majorana neutrino mass matrix, we assume the following democratic mass matrix 
with small violations in all matrix elements except for 33 element 
\begin{equation}
M^{\nu}_M=\Gamma^{\nu}_M\left(\begin{array}{ccc}
                1-\Delta^{\nu}_1&1-\Delta^{\nu}_2&1-\Delta^{\nu}_3\\
                1-\Delta^{\nu}_2&1-\Delta^{\nu}_4&1-\Delta^{\nu}_5\\
                1-\Delta^{\nu}_3&1-\Delta^{\nu}_5&1
                \end{array}\right),\ \   
\Delta^{\nu}_i\ll1\ \ (i=1,2,3,4,5),\ \ \ 
\end{equation}
adding breaking terms to the (1,1) and (2,2) elements in order 
to keep the generality. This pattern of violation is most general form on the 
left-right symmetry scenario.
Branco et al.~\cite{BRANCO} assumed the pattern of violations similar to their 
charged lepton and Dirac neutrino mass matrices for Majorana neutrinos 
$M^\nu_M$. 
\begin{equation}
M^{\nu}_M=\Gamma^{\nu}_M\left(\begin{array}{ccc}
                1&1&1\\
                1&1+\varepsilon^{M}_1&1\\
                1&1&1+\varepsilon^{M}_2
                \end{array}\right),\ \ \varepsilon^M_i\ll1\ \  (i=1,\ 2).\ \ \ 
\end{equation}
\par
The charged lepton mass matrix $M^l$ in Eq.~(1) is diagonalized by the unitary 
matrix $U^l$ close to $U_0$ in Eq.~(4). Thus if the effective left-handed 
Majorana neutrino mass matrix $M_{\rm eff}^{\nu}$ is diagonalized by the 
unitary matrix $U^{\nu}$ close to unit matrix, the neutrino mixing matrix 
$M_{\rm MNS}$ defined by $U_{\rm MNS}=U^l{U^{\nu}}^\dagger$ is close to the 
unitary matrix $U_0$ as 
\begin{equation}
U_{\rm MNS}\sim U_0.
\end{equation}
This result gives the large $e$-$\mu$ and $\mu$-$\tau$ mixing, thus the 
neutrino bimaximal mixing is induced from seesaw mechanism. Main purpose of 
this work is to show that one can obtain the allowed values of violation 
parameters satisfying the condition that the unitary matrix $U^{\nu}$ 
diagonalizing the effective neutrino mass matrix $M^{\nu}_{\rm eff}$ becomes 
nearly unit matrix.
%%%%%%%%%%%%%%%%%%%%% Section 3 %%%%%%%%%%%%%%%%%%%%%%%%%   
%%%%%%%%%%%%%%%%%%%%% Section 3-1 %%%%%%%%%%%%%%%%%%%%%%%            
\section{Violation parameters of Dirac and Majorana neutrino mass matrix }
\subsection{Neutrino Large Mixing induced through Seesaw Mechanism}
First, we study a general property of the $M_{\rm eff}=M_DM^{-1}_MM^T_D$ 
concerned with violation parameters in $M_D$ and $M_M$, where we suppressed 
the superscript $\nu$ in $M_D$ and $M_M$. We represent the mass 
matrix $M_D$ and $M_M$ as
\begin{equation}
M_D=\Gamma_D(\Sigma+P_D),\ \ \ \ M_M=\Gamma_M(\Sigma+P_M),
\end{equation}
where
\begin{equation}
\Sigma=\left(\begin{array}{ccc}
                1&1&1\\
                1&1&1\\
                1&1&1
                \end{array}\right),
\end{equation}
and $P_D$ and $P_M$ are matrices of violations as  
\begin{subequations}
\begin{equation}
P_D=\left(\begin{array}{ccc}
                0&-\delta_1&-\delta_2\\
                -\delta_1&0&-\delta_3\\
                -\delta_2&-\delta_3&0
                \end{array}\right),\ \ \ 
P_M=\left(\begin{array}{ccc}
                -\Delta_1&-\Delta_2&-\Delta_3\\
                -\Delta_2&-\Delta_4&-\Delta_5\\
                -\Delta_3&-\Delta_5&0
                \end{array}\right),
\end{equation}
for our analysis, and  
\begin{equation}
P_D=\left(\begin{array}{ccc}
                0&0&0\\
                0&\varepsilon_1^{D}&0\\
                0&0&\varepsilon_2^{D}
                \end{array}\right),\ \ \ 
P_M=\left(\begin{array}{ccc}
                0&0&0\\
                0&\varepsilon_1^{M}&0\\
                0&0&\varepsilon_2^{M}
                \end{array}\right),
\end{equation}
\end{subequations}  
for Branco's analysis \cite{BRANCO}.  
\par
Inverse matrix of $M_M$ in $M^{\rm eff}$ is 
\begin{equation}
M_M^{-1}=\Gamma_M^{-1}(\Sigma+P_M)^{-1}=\frac{1}{\Gamma_M}
\frac{1}{D_2+D_3}(L+Q),
\end{equation}
where $D_2$ and $D_3$ are quadratic and cubic polynomials of 
violation parameters $\Delta_i$ or $\varepsilon^M_i$ in determinant of matrix 
$\Sigma+P_M$, respectively. $L$ and $Q$ are matrices with linear and quadratic 
elements of violation parameters $\Delta_i$ or $\varepsilon^M_i$ in $P_M$. 
Using the relations as shown in literature 
\cite{BRANCO}, 
\begin{equation}
\Sigma L=L\Sigma=0,\ \ \ \sum_{i, j}Q_{ij}=D_2,\ \ \ \Sigma Q \Sigma=D_2\Sigma,  
\end{equation}
we can obtain a general expression for $M_{\rm eff}$
\begin{equation}
M_{\rm eff}=\frac{\Gamma_D^2}{\Gamma_M}\frac{1}{D_2+D_3}
\left[D_2\Sigma+P_DLP_D+\Sigma QP_D+P_DQ\Sigma+P_DQP_D\right].
\end{equation}
In Eq.(17), the first term is quadratic in $\Delta_i$ or $\varepsilon^M_i$ and 
the second term is linear in $\Delta_i$ or $\varepsilon^M_i$ and quadratic in 
$\delta_i$ or $\varepsilon^D_i$, and the third, forth and  fifth terms are 
quadratic in $\Delta_i$ or $\varepsilon^M_i$ times linear, linear and 
quadratic in $\delta_i$ or $\varepsilon^D_i$, respectively. Thus, if the 
smallness of the small violation parameters $\Delta_i$ or $\varepsilon^M_i$ 
and $\delta_i$ or $\varepsilon^D_i$ are same order, the first term is 
dominant over other terms. In this case, $M_{\rm eff}$ is close on the 
democratic type mass matrix, and then $U^{\nu}$ is close on $U_0$ and 
$U_{\rm MNS}$ cannot give the large mixing. However, if small violation 
parameters $\delta_i$ and $\Delta_i$ or $\varepsilon_i^{M}$ and 
$\varepsilon_i^{D}$ are satisfy the 
relation 
\begin{equation}
\Delta_i\ll\delta_i\ll1,\ \ \ \varepsilon_i^{M}\ll\varepsilon_i^{D},
\end{equation}
the second term is dominant over other terms. In this case, $M_{\rm eff}$ 
can be apart from the democratic mass matrix and can produce the unitary 
matrix diagonalizing the $M_{\rm eff}$ close to the unit matrix. Thus we can 
induce the neutrino bi-maximal mixing through the seesaw mechanism adopting 
the violation parameters satisfying Eq. (18).
%%%%%%%%%%%%%%%%%%%% Section3-2 %%%%%%%%%%%%%%%%%%%%%%%%%%%%%%%
\subsection{Numerical Result}
Here, we would like to each allowed numerical values of violation parameters 
in Dirac and Majorana neutrino mass matrices producing the observed neutrino 
mass hierarchy and large neutrino mixing. Solar neutrino mass and atmospheric 
neutrino mass are observed as $\Delta m^2_{\odot}\sim10^{-5}{-}10^{-4}
~{\rm eV}^2$ and $\Delta m_{\rm atm}^2\sim(1.5-5)\times10^{-5}~{\rm eV}^2$, 
respectively, then 
\begin{equation}
\frac{m_{\nu_e}}{m_{\nu_\mu}}<0.34,\ \ \ \ 
0.03<\frac{m_{\nu_\mu}}{m_{\nu_\tau}}<0.3.
\end{equation} 
For $U^{\nu}$, we set the off diagonal matrix elements smaller than 0.15, 
\begin{equation}
{\rm off\ diagonal\ element\ of\ }U^{\nu}\le 0.15.
\end{equation}
In our previous analysis \cite{TESHIMA1}, we searched the allowed numerical 
values of violation parameters fixing a condition $D_2=0$ in Eq.~(17) in 
order to make a democratic part $D_2\Sigma$ in Eq.~(17) not affect. In present 
analysis, we would not impose such condition.
%%%%%%%%%%%%%Subsubsection%%%%%%%%%%%%%%%%%% 
\subsubsection{Non-diagonal Violation Parameter Case}
First, we analyze the Dirac and Majorana mass matrices with non-diagonal 
violation parameters as 
$$
P_D=\left(\begin{array}{ccc}
                0&-\delta_1&-\delta_2\\
                -\delta_1&0&-\delta_3\\
                -\delta_2&-\delta_3&0
                \end{array}\right),\ \ \ 
P_M=\left(\begin{array}{ccc}
                -\Delta_1&-\Delta_2&-\Delta_3\\
                -\Delta_2&-\Delta_4&-\Delta_5\\
                -\Delta_3&-\Delta_5&0
                \end{array}\right).
$$ 
In Fig.~1, we showed the allowed region of violation paramours for 
($\delta_1$, $\delta_2$, $\delta_3$) limiting $-0.2<\delta_3<0.2$. 
%%%%%%%%%%%%%%%%%%% fig.1%%%%%%%%%%%%%%%%%%%%%%
\begin{figure}[htbp]
\includegraphics{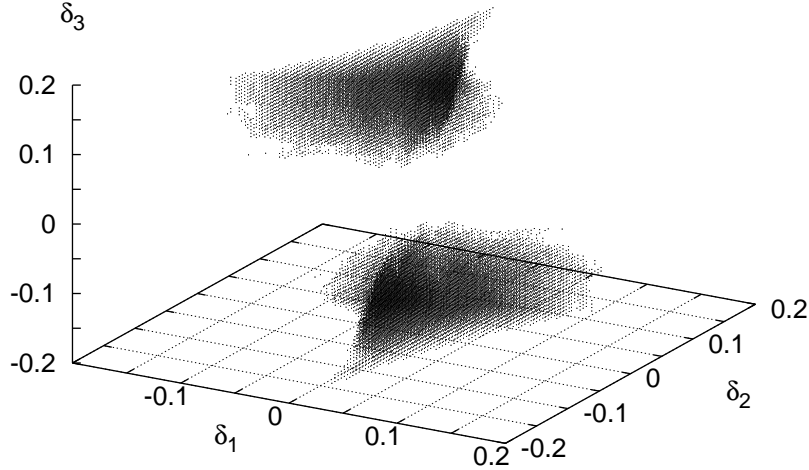}
\vspace{-0.7cm}\\
\caption{Allowed region of violation parameters for ($\delta_1$, 
$\delta_2$, $\delta_3$) limiting $-0.2<\delta_3<0.2$}
\label{fig1}
\end{figure} 
%%%%%%%%%%%%%%%%%%%%%%%%%%%%%%%%%%%%%%%%%%%%%%%
In order to read the allowed values distinctly, we showed the allowed region 
of ($\delta_1$, $\delta_2$) fixing $\delta_3=0.2,\ 0.15,\ 0.1,\ 0.05$ in 
Fig.2 (a), (b), (c), (d), respectively. For negative $\delta_3$, allowed 
values of $\delta_1$ and  $\delta_2$ are exchanged to opposite sign. 
%%%%%%%%%%%%%%%%%%% fig.2%%%%%%%%%%%%%%%%%%%%%%   
\begin{figure}[htbp]
\includegraphics[width=7.5cm,]{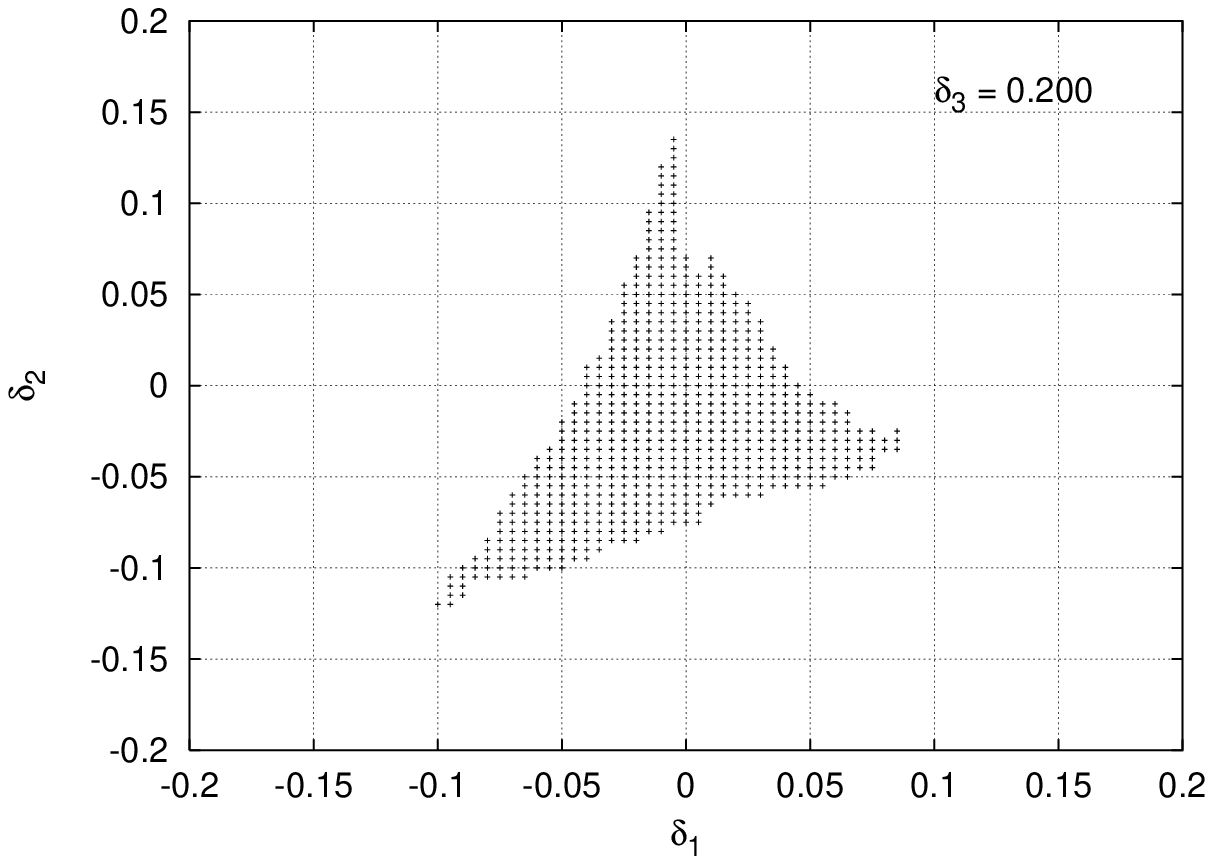}
\includegraphics[width=7.5cm,]{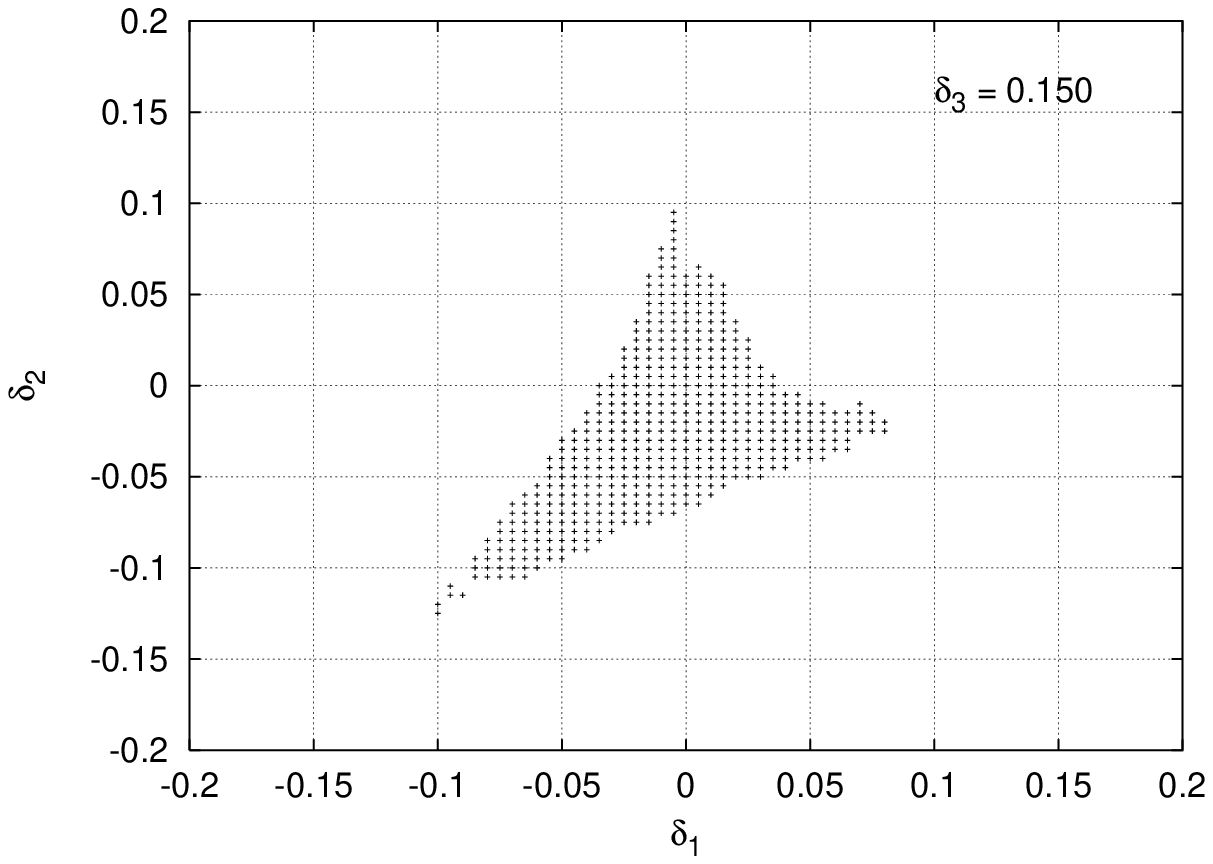}
\vspace{-0.5cm}\\
\center{(a)\hspace{7.5cm} (b)}
\vspace{0.5cm}\\
\includegraphics[width=7.5cm,]{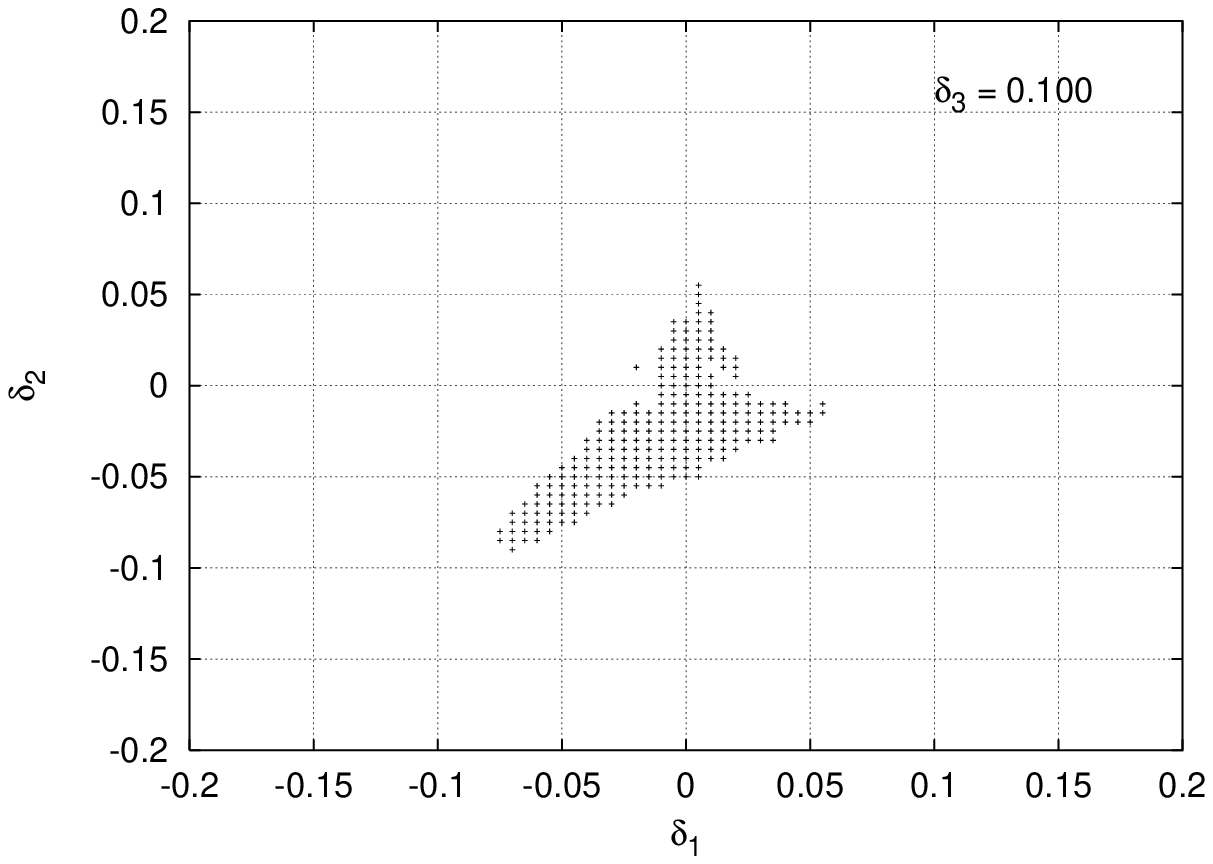}
\includegraphics[width=7.5cm,]{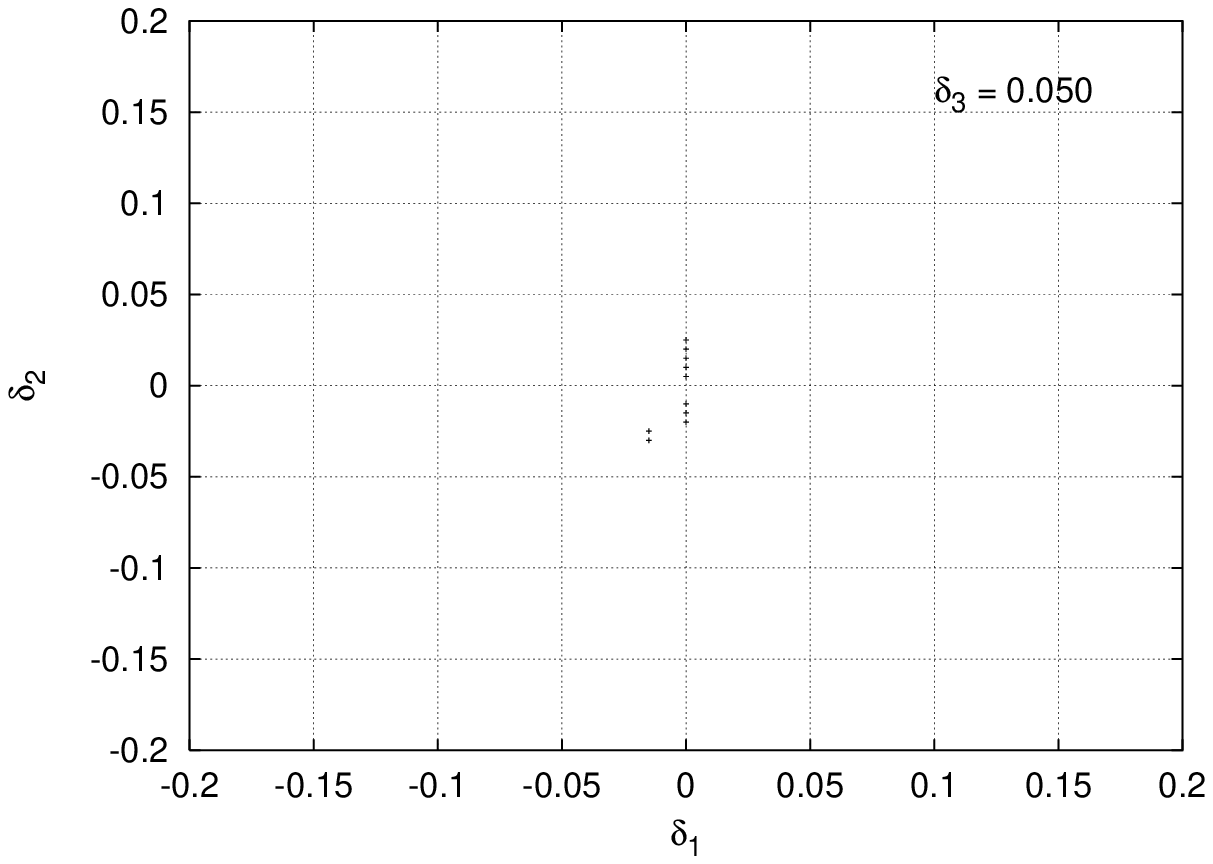}
\vspace{-0.5cm}\\
\center{(c)\hspace{7.5cm} (d)}
\vspace{0.0cm}\\
\caption{Allowed region of ($\delta_1$, $\delta_2$) 
fixing $\delta_3=0.2,\ 0.15,\ 0.1,\ 0.05$ }
\label{fig2}
\end{figure} 
%%%%%%%%%%%%%%%%%%%%%%%%%%%%%%%%%%%%%%%%%%%%%%%
\par
We showed the allowed region of violation parameters for $(\Delta_2, 
\Delta_{14}, \Delta_{35})$ where $\Delta_{14}=\Delta_1+\Delta_4$ and 
$\Delta_{35}=\Delta_3+\Delta_5$ in Fig.~3.
%%%%%%%%%%%%%%%%%%% fig.3%%%%%%%%%%%%%%%%%%%%%%
\begin{figure}
\includegraphics[width=12cm]{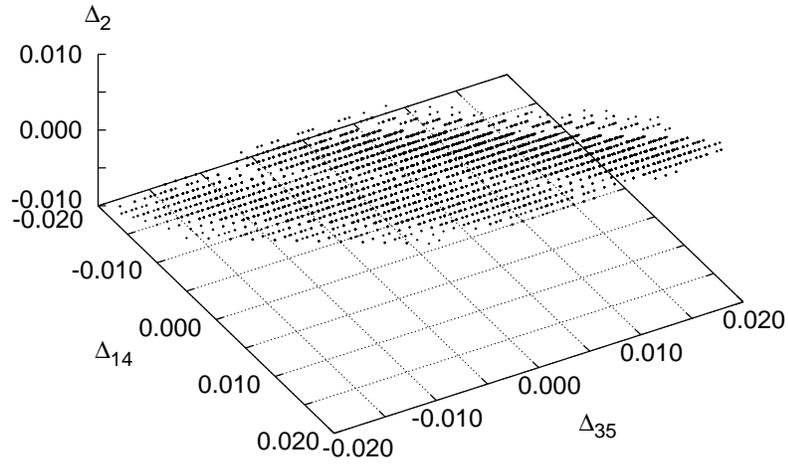}
\vspace{-0.7cm}\\
\caption{Allowed region of violation parameters for $(\Delta_2, 
\Delta_{14}, \Delta_{35})$. $\Delta_{14}=\Delta_1+\Delta_4$, 
$\Delta_{35}=\Delta_3+\Delta_5$. }
\label{fig3}
\end{figure}
%%%%%%%%%%%%%%%%%%%%%%%%%%%%%%%%%%%%%%%%%%%%%%%%% 
In Fig.~4, projected regions of allowed $(\Delta_2, \Delta_{14}, \Delta_{35})$ 
to $(\Delta_{14}, \Delta_{35})$, $(\Delta_2, \Delta_{14})$ and 
$(\Delta_2, \Delta_{35})$ planes are shown. 
\par
From this result, we can recognize that 
(1) $\Delta_i$'s are smaller than $\delta_i$'s as discussed in previous 
subsection, 
(2) numerical values in rather large range of violation parameters $\delta_i$'s 
can produce the neutrino large mixing, 
(3) $|\delta_3|$ is not smaller than 0.05 and $\delta_1$ and $\delta_2$ are 
mutually symmetric, 
(4) the values of violation parameters $\Delta_i$'s of the Majorana neutrinos 
are rather restricted compared to the violation parameters of $\delta_i$' of 
Dirac neutrinos, 
(5) in parameters $\Delta_i$'s, $\Delta_{14}$ and $\Delta_{35}$ 
seem independent parameters rather than $\Delta_1$, $\Delta_3$, 
$\Delta_4$ and $\Delta_5$. 
%%%%%%%%%%%%%%%%%%% fig.4%%%%%%%%%%%%%%%%%%%%%%
\begin{figure}
\includegraphics[width=7.5cm]{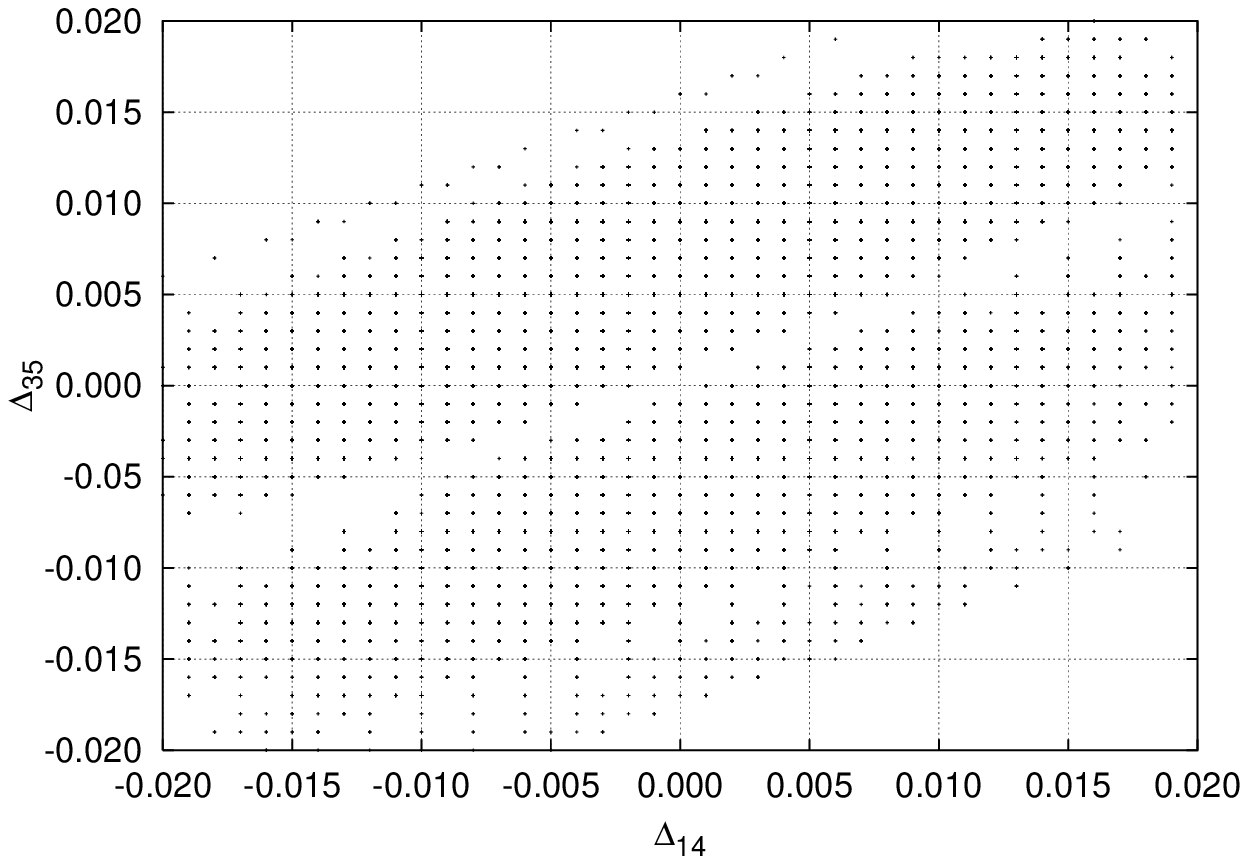}
\includegraphics[width=7.5cm]{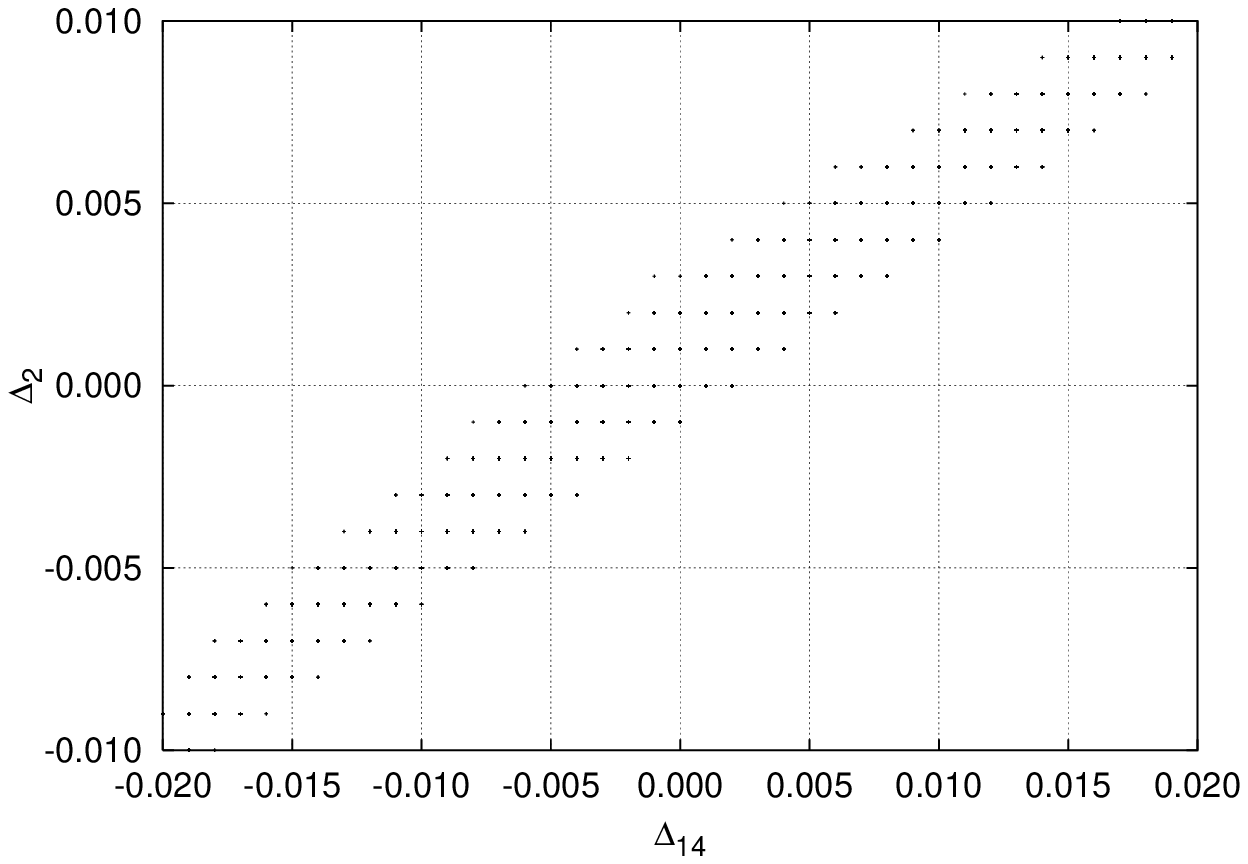}
\vspace{-0.5cm}\\
\center{(a)\hspace{7.3cm}(b)}
\vspace{0.3cm}\\
\includegraphics[width=7.5cm]{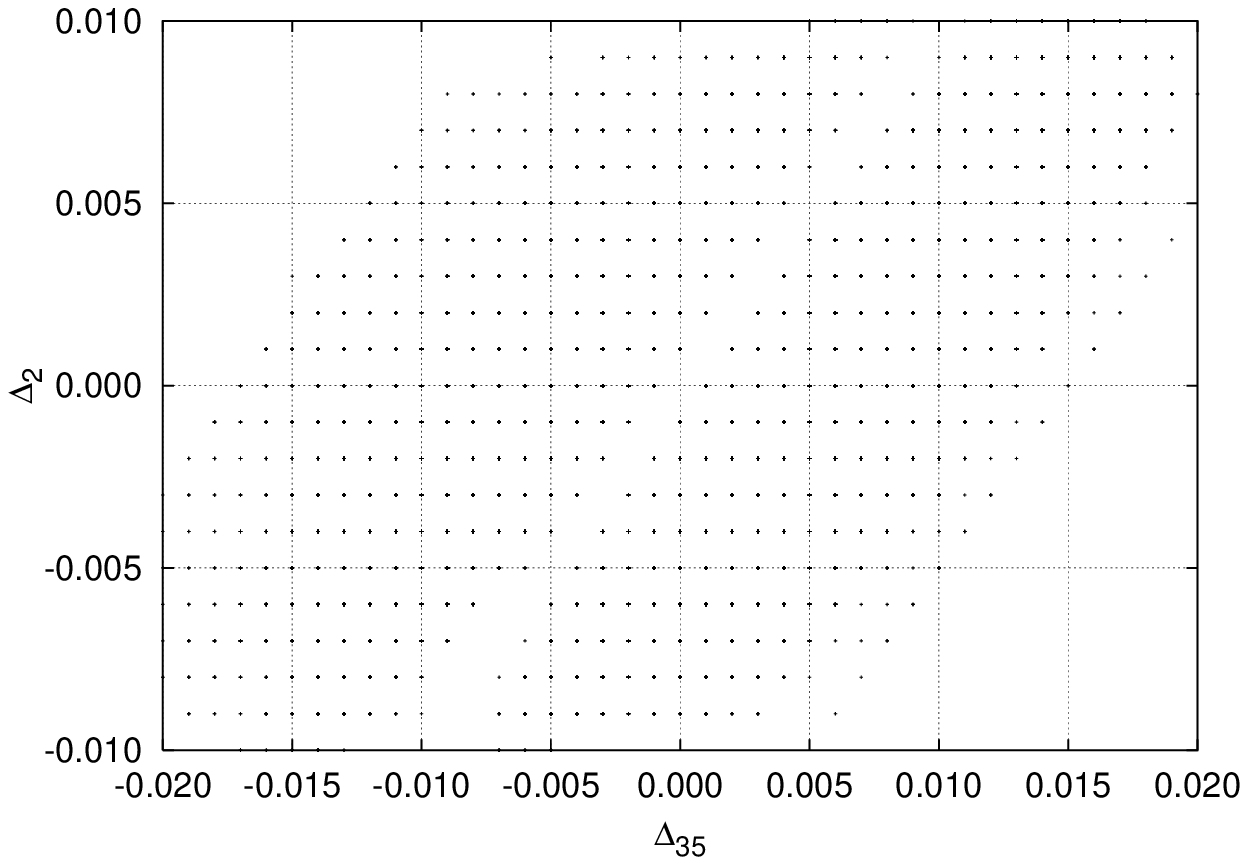}\hspace{7.6cm}
\vspace{-0.5cm}\\
\center{(c)\hspace{7.6cm}}
\vspace{-0.2cm}\\
\caption{Projected regions of allowed $(\Delta_2, \Delta_{14}, \Delta_{35})$ 
to $(\Delta_{14}, \Delta_{35})$, $(\Delta_2, \Delta_{14})$ and 
$(\Delta_2, \Delta_{35})$ planes}
\label{fig4}
\end{figure}
%%%%%%%%%%%%%%Subsubsection%%%%%%%%%%%%%%%%%
\subsubsection{Diagonal Violation Parameter Case} 
For this case, Branco {\it et al.} discussed precisely \cite{BRANCO}. They 
used the Dirac and Majorana Mass matrices with diagonal violation parameters 
as   
$$
P_D=\left(\begin{array}{ccc}
                0&0&0\\
                0&\varepsilon_1^{D}&0\\
                0&0&\varepsilon_2^{D}
                \end{array}\right),\ \ \ 
P_M=\left(\begin{array}{ccc}
                0&0&0\\
                0&\varepsilon_1^{M}&0\\
                0&0&\varepsilon_2^{M}
                \end{array}\right).
$$   
In this case, $M_{\rm eff}$ is written as 
\begin{equation}
M_{\rm eff}=\frac{\Gamma_D^2}{\Gamma_M}
\left(\begin{array}{ccc}
          1&1&1\\
          1&1+\frac{{\varepsilon^D_1}^2}{\varepsilon_1^M}&1\\
          1&1&1+\frac{{\varepsilon^D_2}^2}{\varepsilon_2^M}
       \end{array}
\right).
\end{equation}
Then if $\frac{{\varepsilon^D_1}^2}{\varepsilon_1^M}$ and 
$\frac{{\varepsilon^D_2}^2}{\varepsilon_2^M}$ are sufficiently larger than 1, 
one can get the large mixing of $U_{\rm MNS}$. Numerically, allowed regions 
of $\varepsilon^D_1$, $\varepsilon_1^M$, $\varepsilon^D_2$ and 
$\varepsilon_2^M$ satisfying the condition Eqs. (19) and (20) are restricted 
as follows;
\begin{subequations}
\begin{eqnarray}
  &&\varepsilon^D_1 = 0.02 \sim 0.04/-0.02 \sim -0.04,\ \ 
  \varepsilon^D_2 = 0.2(\rm fixed),\nonumber\\
  &&\hspace{1cm}\varepsilon_1^M= 0.000001 \sim 0.00025/-0.000001 \sim 
  -0.00025,\nonumber\\ 
  &&\hspace{1cm}\varepsilon_2^M = 0.00001 \sim 0.0016/-0.00001 \sim 
  -0.0018,\\
  &&\varepsilon^D_1 = 0.01 \sim 0.02/-0.01 \sim -0.02,\ \  
  \varepsilon^D_2 =  0.1(\rm fixed),\nonumber\\
  &&\hspace{1cm}\varepsilon_1^M= 0.000007 \sim 0.000063/-0.000007 \sim 
  -0.000063, \nonumber\\ 
  &&\hspace{1cm}\varepsilon_2^M = 0.000075 \sim 0.000375/-0.000075 \sim 
  -0.00045, \\
  &&\varepsilon^D_1= 0.005 \sim 0.01/-0.005 \sim -0.01,\ \  
  \varepsilon^D_2 = 0.05(\rm fixed),\nonumber\\
  &&\hspace{1cm}\varepsilon_1^M= 0.000002 \sim 0.000016/-0.000002 \sim 
  -0.000016, \nonumber\\ 
  &&\hspace{1cm}\varepsilon_2^M= 0.000015 \sim 0.000105/-0.000015 \sim 
  -0.000105.
\end{eqnarray}
\end{subequations}
$\varepsilon^D_2$ can take arbitrary value, then we showed the cases
fixed to 0.2, 0.1 and 0.05. Furthermore in the above 
allowed regions , we restricted the $\varepsilon^D_1$ to the ranges where 
the ratio $\varepsilon^D_1/\varepsilon^D_2$ is smaller than 0.2. From this 
result, we can say that (1) $\varepsilon^M_i$ are smaller than 
$\varepsilon^D_i$ as discussed in previous subsection, (2) numerical values 
in rather large range of violation parameters $\varepsilon^D_i$'s 
can produce the neutrino large mixing.
%%%%%%%%%%%%%Subsubsection%%%%%%%%%%%%%%%%%
\subsubsection{Comparison of the Violating Patterns on the Quark Sector 
and the Neutrino Sector}
We present the case  analyzed precisely in our work \cite{TESHIMA2} 
where violation parameters are involved in nondiagonal element of the quark 
mass matrix. We use the following quark mass matrices Eq.(3) with small 
phases:
\begin{eqnarray}
&M_q=\Gamma_q&\left(\begin{array}{ccc}
                1&(1-\delta^q_1)e^{i\phi^q_1}&(1-\delta^q_2)e^{i\phi^q_2}\\
                (1-\delta^q_1)e^{-i\phi^q_1}&1&(1-\delta^q_3)e^{i\phi^q_3}\\
                (1-\delta^q_2)e^{-i\phi^q_2}&(1-\delta^q_3)e^{-i\phi^q_3}&1
                \end{array}\right),\ \ (q=u,\ d)\nonumber\\ 
&&\delta^q_i\ll1,\ \  \phi^q_i\ll1, \ \ (i=1,\ 2,\ 3). 
\end{eqnarray}
Diagonalising these matrices and fitting these eigenvalues to the numerical 
mass ratios and the CKM matrix $U_{\rm CKM}$ to the observed values,
\begin{eqnarray}
&&\frac{m_u}{m_c}=0.0038\pm0.0025,\ \ \ \frac{m_d}{m_s}=0.050\pm0.035,
          \nonumber\\
&&\frac{m_c}{m_t}=0.0042\pm0.0013,\ \ \ \frac{m_s}{m_b}=0.038\pm0.019,
         \nonumber\\
&&V_{\rm CKM}=\left(\begin{array}{ccc}
          0.9747\mbox{--}0.9759&0.218\mbox{--}0.224&0.002\mbox{--}0.005\\
          0.218\mbox{--}0.224&0.9738\mbox{--}0.9752&0.032\mbox{--}0.048\\
          0.004\mbox{--}0.015&0.030\mbox{--}0.048&0.9988\mbox{--}0.9995
          \end{array}\right),
\end{eqnarray}
we were able to obtain a result for the violation parameters:
\begin{eqnarray}
&&\delta^u_1=0.00001\mbox{--}0.0004,\ \ \delta^u_+\equiv\frac{\delta^u_2+
           \delta^u_3}2=0.0064\mbox{--}0.0125,\ \ \ \delta^u_-\equiv
           \delta^u_2-\delta^u_3=\pm(0.0\mbox{--}0.0043),\nonumber\\
&&\delta^d_1=0.001\mbox{--}0.015,\ \ \delta^d_+\equiv\frac{\delta^d_2+
           \delta^d_3}2=0.040\mbox{--}0.129,\ \ \ \delta^d_-\equiv
           \delta^d_2-\delta^d_3=\pm(-0.038\mbox{--}-0.006),\nonumber\\
&&\hspace*{3cm}\phi^d_+\equiv\frac{\phi^d_2+\phi^d_3}2=-4^{\circ}\mbox{--}
           -3^{\circ},\ \ \ \phi^d_-\equiv\phi^d_2-\phi^d_3=\pm(
           -1^{\circ}\mbox{--}0^{\circ}).
\end{eqnarray}
Using these values for the parameters, we obtained the unitarity triangle 
parameters $\rho$ and $\eta$ characterizing the {\it CP} violation. The values 
for $\rho$ and $\eta$ are in close agreement with the results obtained from 
experiment (see Ref. \cite{TESHIMA2}). These parameters display a 
power law behavior parameterized by only 2 parameters, $\lambda$ and $\phi$, 
as 
\begin{eqnarray}
&&\delta^u_1=\lambda^8, \ \ \delta^u_-=\lambda^6,\ \ \delta^u_+=
           \lambda^4,\ \ \delta^d_1=\lambda^4, \ \ \delta^d_-=\lambda^3,
           \ \ \delta^d_+=\lambda^2,\nonumber\\
&&\lambda\approx0.32,\ \ \ \phi_+\equiv\phi\approx-4^\circ.
\end{eqnarray} 
Violation parameters in quark sector have a remarkable character as power 
rule, and then the character should tell a hint for the flavor symmetry. 
We can take the allowed regions of the violation parameter in 
neutrino sector, but cannot take a remarkable character as quark sector, 
because one does not have the precise data of $U_{\rm MNS}$ and mass of 
neutrinos at present. In order to obtain the more restrictive character for 
the violation parameters in neutrino sector, one has to take the more precise 
data on the neutrino experiments, for example mass values of neutrinos and 
mixing parameters of $U_{\rm MNS}$ and CP violation phases.   
%%%%%%%%%%%%%% section 4 %%%%%%%%%%%%%%%%%%%%%%%%
\section{Discussions}
We have discussed the possibility that the universal Yukawa coupling 
(democratic mass matrix) with small violations of Dirac and Majorana neutrinos 
can induce the large mixing of neutrinos through the seesaw mechanism. That 
possibility can be achieved by the condition that the violation parameters of 
Majorana neutrinos are sufficiently smaller than the violation parameters of 
Dirac neutrinos. And then, we analyzed the numerical condition for the 
violation parameters in which the observed mass hierarchy of neutrinos and 
bi-maximal neutrino mixing is satisfied by the violation parameters. We gave 
the allowed numerical results of violation parameters in both cases where 
violation parameters are involved in nondiagonal elements of Dirac and 
Majorana mass matrices and in diagonal elements of those.
\par
Obtained result for violation parameters has following characters: 
Numerical values in rather large range of violation parameters $\delta_i$'s 
can produce the neutrino large mixing. $|\delta_3|$ is not smaller than 0.05 
and $\delta_1$ and $\delta_2$ are mutually symmetric. 
The values of violation parameters of the Majorana neutrinos are rather 
restricted compared to the parameters of Dirac neutrinos. 
$\Delta_{14}$ and $\Delta_{35}$ seem independent parameters rather than 
$\Delta_1$, $\Delta_3$, $\Delta_4$ and $\Delta_5$ in non-diagonal violation 
case. The violation parameters of neutrino mass matrices are not so 
restricted in contrast to the quark sector. 
In order to obtain the distinctive characters for violation parameters in 
neutrino sector, one should take the more precise data concerned with the 
neutrino experiments for example  mass values of neutrinos and 
mixing parameters of $U_{\rm MNS}$ and CP violation phases.  
%%%%%%%% references %%%%%%%%%%%%%%%%%%%%%%%%%

\end{document}